\documentclass[conference,a4paper]{IEEEtran}
\IEEEoverridecommandlockouts

\usepackage[hidelinks]{hyperref}
\usepackage[cmex10]{amsmath}
\usepackage{amssymb,amsfonts}
\usepackage{dblfloatfix}

\usepackage[ruled,vlined]{algorithm2e}
\usepackage{graphicx}
\graphicspath{{Figures/PDF/}{Figures/PNG/}}

\usepackage{booktabs}
\usepackage{siunitx}

\usepackage{texnames}
\usepackage{bm,bbm}
\usepackage{orcidlink}

\usepackage{multirow}
\usepackage{subcaption}
\usepackage[numbers,sort&compress]{natbib}
\usepackage{lineno}

\usepackage{draftwatermark}  
\SetWatermarkText{Preprint}  
\SetWatermarkScale{1.0}      
\SetWatermarkColor[gray]{0.96} 

\begin{document}

\title{\uppercase{Depth Separable architecture for Sentinel-5P Super-Resolution}
}

\author{
	\IEEEauthorblockN{Hyam Omar Ali$^{a,b}$\orcidlink{0000-0002-6927-1525}, Romain Abraham$^{a}$, Bruno Galerne$^{a,c}$\orcidlink{0000-0002-2936-6247}%
	}
	\IEEEauthorblockA{
		$^a$ Institut Denis Poisson,  Universit\'e d'Orl\'eans, Université de Tours, CNRS, France%
	}
	\IEEEauthorblockA{
	$^b$ Faculty of Mathematical Sciences, University of Khartoum, Sudan
	}
	\IEEEauthorblockA{
		$^c$ Institut universitaire de France (IUF)
	}
}

\maketitle
\begin{abstract}
Sentinel-5P (S5P) satellite provides atmospheric measurements for air quality and climate monitoring. While the S5P satellite offers rich spectral resolution, it inherits physical limitations that restricts its spatial resolution. Super-resolution (SR) techniques can overcome these limitations and enhance the spatial resolution of S5P data. In this work, we introduce a novel SR model specifically designed for S5P data that have eight spectral bands with around 500 channels for each band. Our proposed S5-DSCR model relies on Depth Separable Convolution (DSC) architecture to effectively perform spatial SR by exploiting cross-channel correlations. Quantitative evaluation demonstrates that our model outperforms existing methods for the majority of the spectral bands. This work highlights the potential of leveraging DSC architecture to address the challenges of hyperspectral SR. Our model allows for capturing fine details necessary for precise analysis and paves the way for advancements in air quality monitoring as well as remote sensing applications.
\end{abstract}


\section{Introduction}
In 2017, the Sentinel-5 Precursor (Sentinel-5P or S5P) satellite was launched by the European Space Agency (ESA) in collaboration with the European Commission aiming for enhanced Earth observation capabilities \cite{s5p_mission}. The satellite monitors the atmospheric species which impact global air quality, greenhouse gases, and ozone layer dynamics \cite{s5p_applications}. It provides data relevant to atmospheric and environmental observations and climate research. This global data supports policy-making and WHO efforts towards improving public health, prioritizing global air quality monitoring and reducing exposure to air pollution \cite{velasco2022update}. 

The  S5P operates into $8$ different spectral bands \cite{s5p_products}. Through these bands, S5P measures radiance data across $3942$ frequency spectral channels and captures hyperspectral (HS) images enabling the spectral analysis. The pixels of the HS images are generated through a combination of scans in the cross-track direction (perpendicular to the satellite's motion) and the ones in the along-track (parallel to the satellite's motion) \cite{TROPOMI_L01b_Specification}. The spatial resolution of the images is approximately $3.5\times5.5\:km^2$ per pixel \cite{s5p_mission, TROPOMI_L01b_Specification}. 

While the S5P satellite offers rich spectral resolution, making it an invaluable tool for environmental monitoring, its spatial resolution is limited by physical constraints, such as the trade-off between pixel size and the amount of light detected, potentially affecting image quality \cite{orych2015review}. These limitations restrict its potential for finer-scale analysis, necessitating advanced image processing techniques such as super-resolution (SR) to enhance spatial resolution and improve the quality of S5P data for more precise environmental monitoring \cite{camps2011remote}.

The unique spectral characteristics and inter-channel relationships of S5P HS data necessitate the development of more specific SR models. The S5Net model has recently been introduced as the first tailored SR approach for S5P data. Initially, this model was designed to handle individual spectral bands and a single channel at a time \cite{carbone2024model}. Later,  S5Net was extended to process multiple spectral bands considering all the channels \cite{carbone2024efficient}.
In this work, we present a novel approach, S5-DSCR, to enhance the spatial resolution of S5P HS images. Our contributions are as follows:

\begin{itemize}
	\item We use Depth Separable Convolution (DSC) architecture that captures both the spatial and spectral relationships within and across channels. This lightweight architecture ensures capturing the interdependencies across all the spectral channels while reducing complexity.
	\item Considering the unique spectral characteristics of each band, we trained the same model separately for each specific band. This approach ensures the model is optimized and efficient for the unique features of each individual band.
	\item We compiled a comprehensive dataset comprising HS images from 15 orbits representing diverse spatial and spectral characteristics and providing maximum coverage of Earth's surface.
\end{itemize}

The plan of this paper is as follows: Section \ref{related_work} reviews the related work. The methodology is described in Section \ref{method} including problem formulation and the proposed method to resolve. Then, Section \ref{experiments_section} details the experimental setup. Section \ref{results_section} presents and analyzes the results. Finally, Section \ref{conclusion} concludes and draws potential perspectives

\section{Related Work}\label{related_work}
SR is an important branch of image processing that produces high-resolution (HR) images from low-resolution (LR) images.  SR techniques are broadly classified into single-image super-resolution (SISR) and multi-image super-resolution (MISR) \cite{su2024review}. Since each S5P scan is unique, this work is focused on SISR approaches.  

Over the years, numerous methods have been proposed to address SISR \cite{su2024review, yang2019deep, ye2023single, bashir2021comprehensive, aburaed2023review}. The classical and earliest approaches are interpolation-based techniques \cite{keys1981cubic, aiazzi2013bi, li2001new,hung2011robust, hung2012fast}, such as nearest neighbour and bicubic interpolation. Variational or statistical techniques \cite{schultz1996extraction, ng2007zoom, fan2017projections,ng2007total, sun2008image}, such as Total Variation (TV), iterative back-projection, and maximum a posteriori (MAP), leverage optimization and regularization techniques to enhance the SISR. However, with the rapid development of resources and techniques, deep learning approaches became dominant and achieved state-of-the-art performance in SR tasks. These approaches map the features of LR images to HR images. Although all the approaches basically employ the same essential ideas, they differ in network architecture and optimization. Designs such as residual networks \cite{xiong2017hscnn, can2018efficient, han2018residual, kaya2019towards, hang2021spectral, nathan2020light, li2022hasic, gewali2019spectral}, dense connections \cite{wang2018esrgan, haris2018deep, zhang2018residual, tong2017image}, attention mechanisms \cite{nathan2020light, li2022hasic, li2020adaptive, li2020hybrid, zheng2021spectral}, GANs \cite{zhu2021semantic, wang2023review, bulat2018learn, zhang2020supervised, wang2018esrgan}, diffusion models \cite{Saharia_etal_SR3_image_super-resolution_via_iterative_refinement_PAMI2021},  and group learning \cite{hang2021spectral, han2018spectral, gewali2019spectral, mei2022learning}, represent the diversity in approaches.

Unlike grey or RGB images, HS images have hundreds of spectral channels. This makes SR more challenging due to the need to map the features of a larger number of channels. The overall complexity of the model increases gradually as the number of channels increases making the process more challenging to consider the spectral dimension \cite{aburaed2023review}. Considering reducing dimensions and applying convolution along the spectral dimension can achieve more improvement in SR for HS images \cite{guarino2023band, qu2021dual, hidalgo2021dimensionality}. However, as HS images exhibit a high correlation along the spectral dimensions, often these methods neglect to account for this correlation \cite{he2023spectral}. DSC addresses this limitation by representing standard convolution across the spectral channels with convolutions that incorporate the features across both the spectral channels and spatial dimension \cite{sifre2014rigidmotion}. Several studies used DSC for SR \cite{hung2019real, hussain2023depth, sun2024enhanced, muhammad2021multi, jiang2020single}. 

Remote sensing imaging, particularly S5P images, differs from classical HS images due to its unique multi-band structure, spanning eight bands and almost 500 spectral channels per band with varying resolutions. For this reason, S5P images require the development of SR models tailored specifically to preserve spectral fidelity and handle its characteristics. The S5Net model introduced the first deep-learning approach designed specifically for S5P images with lightweight architecture \cite{carbone2024model}. Initially, this model was developed for the monochromic images (central spectral channel only), and S5net was later modified to process all the channels of the S5P images through the fine-tuning cascade procedure \cite{carbone2024efficient}. This approach begins at the central channel and iteratively expands to include all the spectral channels. However, it still processes each channel individually, missing the utilization of inter-channel information.

\section{Proposed Method}\label{method}
\subsection{Problem Formulation}
The objective of this study is to enhance the spatial resolution of Sentinel-5P Level-1B radiance products using  SISR techniques. For each band, the HS image consists of multiple spectral channels with relatively low spatial resolution. Following \cite{carbone2024efficient}, to demonstrate the feasibility of SISR in S5P data, the original HS images were degraded to obtain the LR images while the original HS images acted as the HR ground truth.

The original degradation model \cite{carbone2024model} was used to simulate the LR images. This process involves simulating the real-world image acquisition process, such as blurring, to create realistic LR images that closely resemble the actual HS images. The degradation process involved applying a Point Spread Function (PSF) to model the blurring effect. The blurring kernel $K$ is simulated by asymmetric Gaussian function with different standard deviations in the along-track and cross-track directions. For S5P data, the standard deviations vary for each detector \cite{carbone2024model}: $0.37$ and $0.36$ (UV), $0.44$ and $0.74$ (UVIS), $0.45$ and $0.74$ (NIR), and $0.15$ and $0.20$ (SWIR) in the across-track and along-track directions, respectively. A scaling factor of $4$ was utilized to downsample the HR images.

\subsection{Architecture}
\begin{figure}[t]
	\centering
	\begin{subfigure}[b]{\linewidth}
		\centering
		\includegraphics[ width=\linewidth]{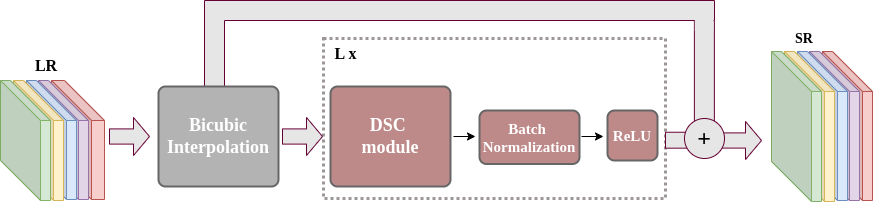}
		\caption{Architecture of S5-DSCR (L=5) and S5-DSCR-S (L=1) models}
		\label{subfig1}
	\end{subfigure}
	\begin{subfigure}[b]{\linewidth}
		\centering
		\includegraphics[height=1.8in, width=\linewidth]{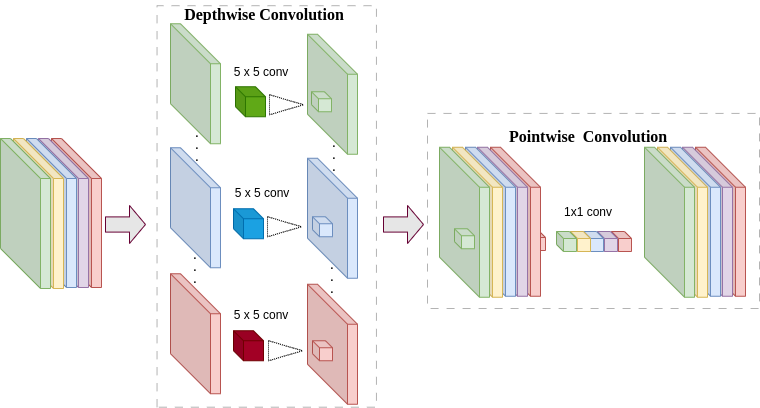}
		\caption{DSC module }
		\label{dsc_module}
	\end{subfigure}
	
	\caption{Overview of our proposed models and their inner DSC module}
	\label{arch}
\end{figure}

Our proposed model is based on the DSC architecture. DSC is composed of two stages (Figure \ref{dsc_module}): depthwise convolution (intra-channel) in which each input channel is independently convolved with a filter to extract features of this channel, followed by a pointwise convolution (intra-pixel) that combines the outputs of the depthwise stage with $1\times1$ convolution creating a new feature map. This representation significantly reduces the number of parameters and computational costs while preserving spectral fidelity \cite{chollet2017xception, junejo2021depthwise}.

We introduce two versions of our model, S5-DSCR and its lightweight variant S5-DSCR-S. Figure \ref{arch} illustrates our proposed model. S5-DSCR incorporates multiple DSC modules with L=5, serving as five recursive layers, and integrates a residual connection to efficiently extract spatial and spectral features from HS images. S5-DSCR-S retains comparable performance with a single DSC module (L=1) making it suitable for resource-constrained environments.

The architecture adopts a cascaded design with a pre-upsampling approach that maintains effective feature extraction. Initially, the LR input image is upsampled using bicubic interpolation to generate an intermediate image that provides a baseline estimation of the final super-resolved (SR) image with the intended spatial dimensions. 
To extract the complex spatial and spectral features, the intermediate image is then given to the DSC module that processes jointly all the channels. 
This module restores missing high-frequency details to be added to the bicubic interpolation.
Each depthwise convolution in the DSC module employs a $5\times5$ kernel, generating the intermediate feature maps. Each of these feature maps has the same size as the input image making the model memory efficient. 
Eventually, the residual connection combines the bicubic interpolation and the DSC output to obtain the final SR image. 

\subsection{Training Details}
To consider the characteristics of each band, the same model was trained separately for each band. This ensures the model is optimized for each band's unique characteristics, such as the number of channels, spatial resolution, noise ratio, etc. Additionally, this approach allows for better adaptation to variations in spectral resolution and spatial complexity. 

We trained both S5-DSCR and S5-DSCR-S. The training was performed on the HS images described below. The data was split into completely separate, non-overlapping training, validation, and testing with 65\%, 20\%, and 15\% respectively, ensuring a balanced representation of spatial and spectral diversity. Each image was divided into overlapping patches of size $64\times64$ and $256\times256$ for LR and HR images, respectively. We employ Mean Square Error (MSE) loss to train the models and optimize them using the Adam optimizer with the default PyTorch parameters. The initial learning rate was set to $10^{-3}$ and reduced by a factor of $0.1$ when the loss value did not improve for three epochs.

\section{Experiments}\label{experiments_section}
\subsection{Dataset}
We used Sentinel-5P radiance data as the primary source of images for this study. These data are freely accessible on Copernicus’ official website \cite{CopernicusDataSpace} as Level-1B radiance data. The dataset comprises data from $15$ orbits covering distinct regions acquired on January 4, 2023, and September 7, 2023. The orbits cover data from diverse geographical regions, including Africa, the Gulf countries, America, Europe, Asia, and New Zealand. These regions were selected to represent various climatic and geographical conditions, ensuring that the dataset captures diverse spatial and spectral characteristics and provides maximum coverage of Earth's surface.
\setlength{\tabcolsep}{2pt} 
\begin{table}[hbt]
	\centering
	\caption{Sentinel-5P spectrometers characteristics}
	\label{s5p info}
	\begin{tabular}{@{}cccc@{}}
		\toprule
		\textbf{Spectrometer} & \textbf{\# Channels} & \textbf{Name} & \textbf{Range (nm)} \\ 
		\midrule
		\multirow{2}{*}{Ultraviolet (UV)} & \multirow{2}{*}{497} & UV-1 / Band 1 & 270-300 \\ 
		& & UV-2 / Band 2 & 300-320 \\
		\multirow{2}{*}{Ultraviolet Visible (UVIS)} & \multirow{2}{*}{497} & UVIS-1 / Band 3 & 320-405 \\ 
		& & UVIS-2 / Band 4 & 405-500 \\
		\multirow{2}{*}{Near-Infrared (NIR)} & \multirow{2}{*}{497} & NIR-1 / Band 5 & 675-725 \\ 
		& & NIR-2 / Band 6 & 725-775 \\
		\multirow{2}{*}{Shortwave-Infrared (SWIR)} & \multirow{2}{*}{480} & SWIR-1 / Band 7 & 2305-2345 \\ 
		& & SWIR-2 / Band 8 & 2345-2385 \\
		\bottomrule
	\end{tabular}
\end{table}

Each orbit contains radiance data of eight distinct bands as indicated in Table \ref{s5p info}. Band 1 was not included in this study because of its low signal-to-noise ratio \cite{carbone2024efficient}. Due to orbital and regional variations, the radiance data exhibits various spatial dimensions. The full radiance image spans a range of $4172$ to $3735$ along-track (scanlines) and $450$ to $215$  across-track (ground pixels), depending on the specific region and band. To ensure consistency and computational efficiency,  we cropped each radiance image into multiple images of size $512\times256$ for all bands and $512\times215$ for SWIR bands. In contrast to \cite{carbone2024model}, which focuses on a region corresponding to the central part of the swath, this cropping allows for the inclusion of more diverse regions. We excluded images exhibiting anomalies or extreme values due to inconsistent radiance measurements or sensor artefacts to further refine the dataset. This was performed using a combination of the interquartile range (IQR) to detect outliers beyond $1.5$ times the IQR and $1\%$ percentile clipping \cite{merry2024backbone, nakanishi2024robust}.

\subsection{Evaluation Metrics}
The performance of the model was evaluated using standard metrics that assess image quality and spectral fidelity. These metrics provide an objective evaluation of the spatial and spectral resolution of each band. We used four measures \cite{wang2004image, an2014orientation, zhang2018unreasonable}: Peak Signal-to-Noise Ratio (PSNR), Spatial Correlation Coefficient (SCC), Structural Similarity Index (SSIM), and Learned Perceptual Image Patch Similarity (LPIPS). PSNR quantifies pixel-level accuracy, and SCC measures spatial feature correlation. Considering the perceptual analysis, SSIM assesses perceptual similarity taking into account luminance, contrast, and structural information while LPIPS evaluates perceptual differences using deep features. These metrics collectively offer a global assessment focusing on both perceptual and structural fidelity.

\section{Results}\label{results_section}
The performance of our proposed models was quantitatively evaluated against the separate test dataset and compared against two methods: cubic interpolation and S5Net \cite{carbone2024efficient}. Since S5Net has three different weights, we used the validation dataset to decide the optimal weight for our data. S5-DSCR has 3.9M/3.6M parameters for (bands 2-6/7-8), respectively, while S5-DSCR-S has 0.25M/0.24M parameters for (bands 2-6/7-8), respectively. In contrast, S5Net has 57k parameters per channel, resulting in 28.5M/27.5M parameters for (bands 2-6/7-8) when applied across all channels. Table \ref{metrics} depicts the evaluation results.


 
\begin{table}[t]
	\centering
	\caption{Performance metrics for SR models on our S5P testing dataset. Best results are in bold, second-best underlined.}
	\label{metrics}
	\begin{tabular}{@{}cccccc@{}}
		\toprule
		\textbf{Band} & \textbf{Method} & \textbf{PSNR $\uparrow$ } & \textbf{SCC $\uparrow$ } & \textbf{SSIM $\uparrow$} & \textbf{LPIPS $\downarrow$ } \\ 
		\midrule
		\multirow{4}{*}{BAND 2} & BICUBIC & 32.13 & 0.846 & 0.940 & 0.107 \\
		& S5NET & 32.30 & 0.845 & \underline{0.934} & \underline{0.058} \\ 
		& S5-DSCR & \textbf{37.16} & \underline{0.866} & \textbf{0.952} & \textbf{0.051} \\
		& S5-DSCR-S & \underline{35.55} & \textbf{0.875} & 0.905 & 0.095 \\
		\midrule
		\multirow{4}{*}{BAND 3} & BICUBIC & 27.10 & 0.887 & 0.875 & 0.194 \\
		& S5NET & 28.98 & 0.846 & 0.906 & 0.135 \\ 
		& S5-DSCR & \textbf{33.80} & \underline{0.910} & \underline{0.915} & \textbf{0.063} \\
		& S5-DSCR-S & \underline{32.74} & \textbf{0.916} & \textbf{0.919} & \underline{0.132} \\
		\midrule
		\multirow{4}{*}{BAND 4} & BICUBIC & 25.43 & 0.772 & 0.776 & 0.231 \\
		& S5NET & 26.28 & \textbf{0.8589} & 0.7467 & 0.2944 \\ 
		& S5-DSCR & \textbf{26.96} & 0.777 & \underline{0.784} & \underline{0.222} \\
		& S5-DSCR-S & \underline{26.83} & \underline{0.788} & \textbf{0.796} & \textbf{0.220} \\
		\midrule
		\multirow{4}{*}{BAND 5} & BICUBIC & 20.21 & 0.683 & 0.609 & 0.387 \\
		& S5NET & \textbf{26.79} & \underline{0.730} & \textbf{0.745} & \textbf{0.269} \\ 
		& S5-DSCR & \underline{24.71} & 0.726 & 0.692 & 0.292 \\
		& S5-DSCR-S & 24.52 & \textbf{0.743} & \underline{0.712} & \underline{0.282} \\
		\midrule
		\multirow{4}{*}{BAND 6} & BICUBIC & 22.62 & 0.757 & 0.710 & 0.273 \\
		& S5NET & \textbf{27.60} & \textbf{0.843} & \textbf{0.759} & 0.269 \\ 
		& S5-DSCR & \underline{25.77} & \underline{0.764} & 0.717 & \underline{0.247} \\
		& S5-DSCR-S & 24.45 & 0.775 & \underline{0.725} & \textbf{0.227} \\
		\midrule
		\multirow{4}{*}{BAND 7} & BICUBIC & 12.54 & 0.794 & 0.523 & 0.320 \\
		& S5NET & 27.05 & 0.818 & \underline{0.854} & 0.446 \\ 
		& S5-DSCR & \textbf{31.94} & \textbf{0.962} & \textbf{0.910} & \textbf{0.0541} \\
		& S5-DSCR-S & \underline{30.02} & \underline{0.867} & 0.811 & \underline{0.195} \\
		\midrule
		\multirow{4}{*}{BAND 8} & BICUBIC & 12.58 & 0.853 & 0.431 & 0.324 \\
		& S5NET & \underline{31.59} & 0.820 & \underline{0.865} & 0.192 \\ 
		& S5-DSCR & \textbf{34.80} & \textbf{0.919} & \textbf{0.896} & \textbf{0.153} \\
		& S5-DSCR-S & 29.91 & \underline{0.913} & 0.819 & \underline{0.183} \\
		\bottomrule
	\end{tabular}
\end{table}
Bicubic interpolation served as a baseline method showing how difficult the SR would be since it is not a learnable method. The heterogeneity in the performance of the bicubic interpolation confirms that each band needs to be processed separately.

The S5-DSCR model consistently demonstrated superior performance across most of the bands. This is indicated by achieving the highest scores in PSNR, SCC, and SSIM while maintaining the lowest LPIPS values. Hence, S5-DSCR has a strong ability to recover HR images with both spatial and spectral fidelity, particularly its ability to maintain significant perceptual quality. On the other hand, S5-DSCR-S captures spectral correlations better than S5-DSCR while maintaining good perceptual quality that is comparable results to S5-DSCR. Notably, S5Net performed reasonably well overall in most of the metrics except for LPIPS, suggesting it has lower perceptual quality. Although bands 7 and 8 possess a low-resolution nature, S5-DSCR demonstrates significant improvements across all metrics. This showcases that  S5-DSCR is robust for challenging bands such as these bands and can address various spatial complexities. On the other hand, the performance on bands 5 and 6 was not superior. We argue that this could be due to the higher variation across channels, especially considering that NIR spectrometers are vulnerable and sensitive to environmental factors such as light and weather conditions \cite{stenberg2010visible}.

\begin{figure}[t]
	\centering
	\begin{subfigure}[b]{\linewidth}
		\centering
		\includegraphics[ width=\linewidth]{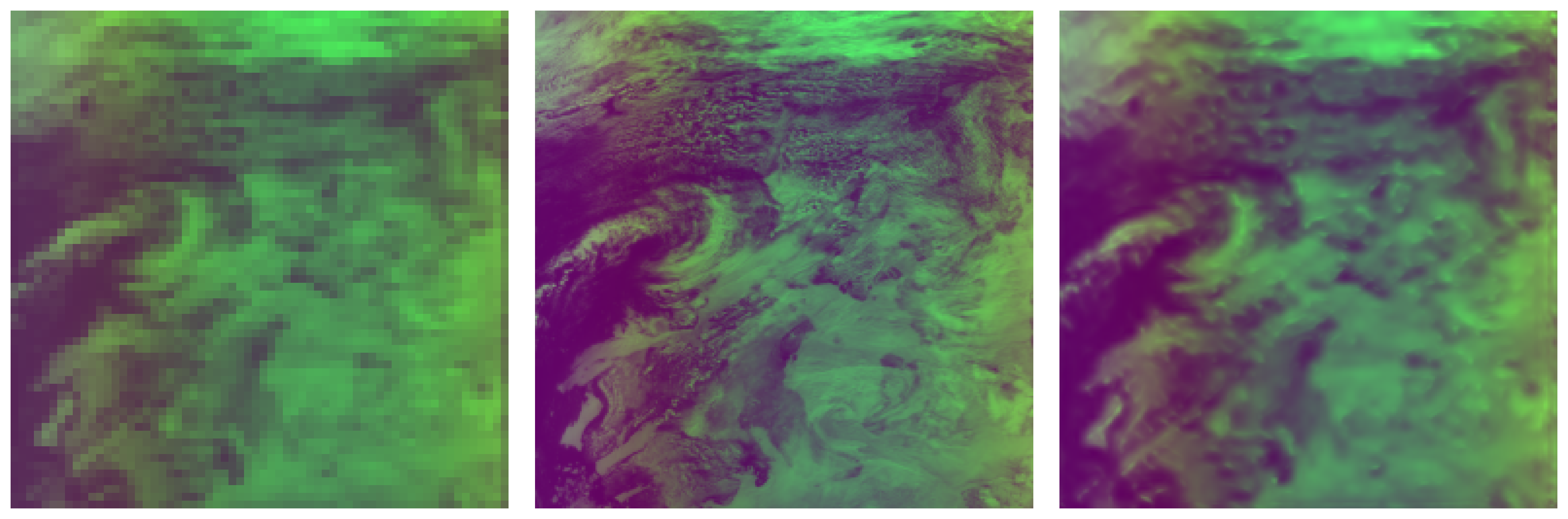}
	\end{subfigure}
	\begin{subfigure}[b]{\linewidth}
		\centering
		\includegraphics[ width=\linewidth]{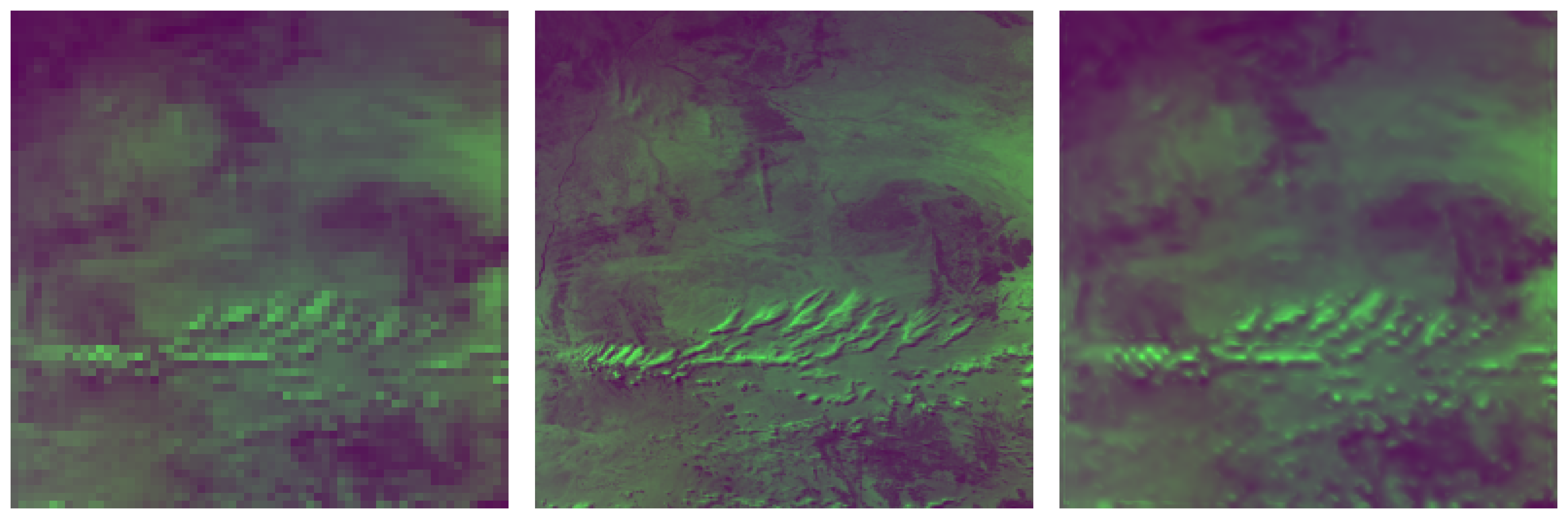}
	\end{subfigure}
	\begin{subfigure}[b]{\linewidth}
		\centering
		\includegraphics[ width=\linewidth]{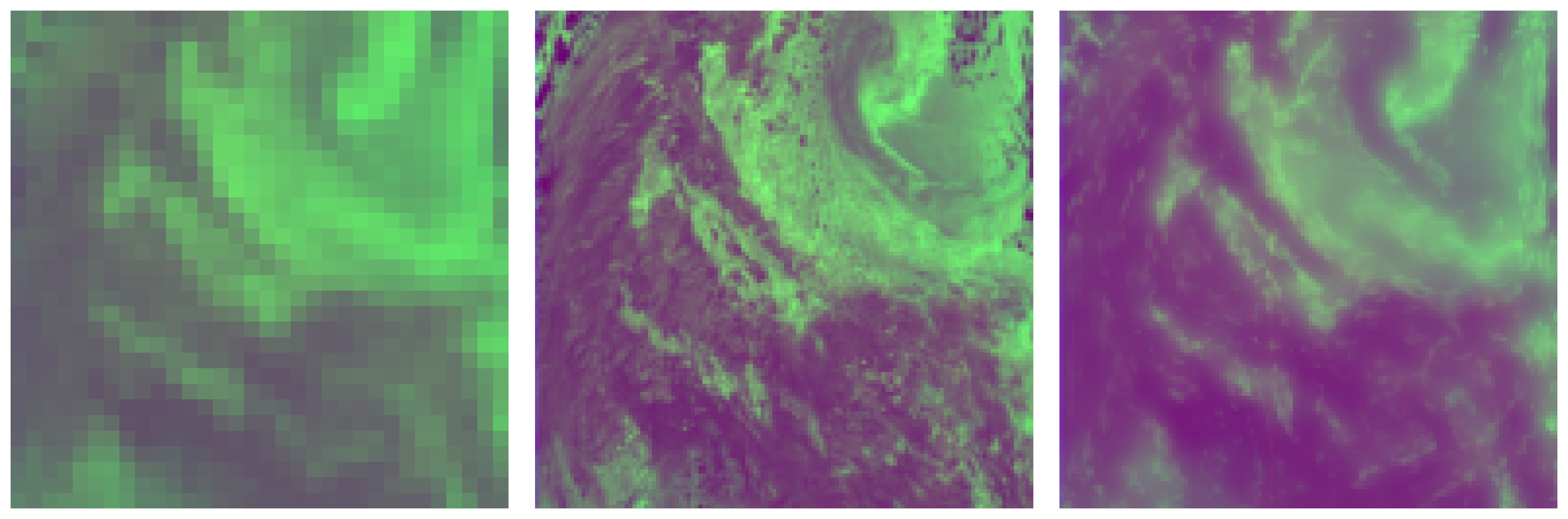}
	\end{subfigure}
	\caption{SR results of S5-DSCR model for, from top to bottom,  bands 3, 5 and 7, with, from left to right, LR, ground truth HR, and our result. For visualization, each image is displayed in the first three PCA components of the ground truth.}
	\label{predication}
\end{figure} 

Additionally, we provide qualitative SR results of the proposed S5-DSCR model in Figure \ref{predication} (only three bands are displayed due to space constraints). For visualization, we applied Principal Component Analysis (PCA) to identify the three most significant components for the channels of the ground truth image. It can be observed that the reconstructed images contain high-frequency details consistent with the ground truth.

Overall, both quantitative and qualitative analyses demonstrate superior performance. Hence, the proposed models are robust and reliable models for S5P data.
 
\section{Conclusion}\label{conclusion}
In this study, we introduced SR models designed specifically for S5P data, utilizing DSC architecture to capture inter-channel correlations while reducing computational complexity. We coupled DSC with residual connections to enhance feature extraction and stability. As a result, our models achieve superior performance across most of the spectral bands while being lightweight and memory efficient. 

A limitation of this work is the reliance on simulated LR data which might not fully provide the degradation patterns. As a future research direction, we plan to explore self-supervised training for the original S5P data as well as testing the model on other satellites datasets to evaluate models' generalization ability.

\vspace{.2cm}
\textbf{Acknowledgment:} The authors acknowledge the financial support of the Junon Project. They are grateful to Alessia Carbone for her invaluable support and insightful discussions. Special thanks also go to Nathalie Moulard and Fabrice Jegou for their valuable contributions to the discussions.



\small
\bibliographystyle{IEEEtranN}
\bibliography{references}

\begin{thebibliography}{65}
\providecommand{\natexlab}[1]{#1}
\providecommand{\url}[1]{#1}
\csname url@samestyle\endcsname
\providecommand{\newblock}{\relax}
\providecommand{\bibinfo}[2]{#2}
\providecommand{\BIBentrySTDinterwordspacing}{\spaceskip=0pt\relax}
\providecommand{\BIBentryALTinterwordstretchfactor}{4}
\providecommand{\BIBentryALTinterwordspacing}{\spaceskip=\fontdimen2\font plus
\BIBentryALTinterwordstretchfactor\fontdimen3\font minus
  \fontdimen4\font\relax}
\providecommand{\BIBforeignlanguage}[2]{{%
\expandafter\ifx\csname l@#1\endcsname\relax
\typeout{** WARNING: IEEEtranN.bst: No hyphenation pattern has been}%
\typeout{** loaded for the language `#1'. Using the pattern for}%
\typeout{** the default language instead.}%
\else
\language=\csname l@#1\endcsname
\fi
#2}}
\providecommand{\BIBdecl}{\relax}
\BIBdecl

\bibitem[{Copernicus Program}(2024{\natexlab{a}})]{s5p_mission}
\BIBentryALTinterwordspacing
{Copernicus Program}, ``{Sentinel-5P Mission Overview},'' 2024, accessed:
  2024-12-23. [Online]. Available:
  \url{https://sentiwiki.copernicus.eu/web/s5p-mission}
\BIBentrySTDinterwordspacing

\bibitem[{Copernicus Programm}(2024)]{s5p_applications}
\BIBentryALTinterwordspacing
{Copernicus Programm}, ``{Sentinel-5P Applications Overview},'' 2024, accessed:
  2024-12-23. [Online]. Available:
  \url{https://sentiwiki.copernicus.eu/web/s5p-applications}
\BIBentrySTDinterwordspacing

\bibitem[Velasco and Jarosi{\'n}ska(2022)]{velasco2022update}
R.~P. Velasco and D.~Jarosi{\'n}ska, ``Update of the who global air quality
  guidelines: Systematic reviews--an introduction,'' \emph{Environment
  International}, vol. 170, p. 107556, 2022.

\bibitem[{Copernicus Program}(2024{\natexlab{b}})]{s5p_products}
\BIBentryALTinterwordspacing
{Copernicus Program}, ``{Sentinel-5P Products Overview},'' 2024, accessed:
  2024-12-23. [Online]. Available:
  \url{https://sentiwiki.copernicus.eu/web/s5p-products}
\BIBentrySTDinterwordspacing

\bibitem[{European Space Agency (ESA)}(2018)]{TROPOMI_L01b_Specification}
\BIBentryALTinterwordspacing
{European Space Agency (ESA)}, \emph{Input/Output Data Specification for the
  TROPOMI L01b Data Processor}, European Space Agency (ESA), 2018, accessed:
  2024-12-21. [Online]. Available:
  \url{https://sentinels.copernicus.eu/documents/247904/3119978/Sentinel-5P-Level-01B-input-output-data-specification}
\BIBentrySTDinterwordspacing

\bibitem[Orych(2015)]{orych2015review}
A.~Orych, ``Review of methods for determining the spatial resolution of uav
  sensors,'' \emph{The International Archives of the Photogrammetry, Remote
  Sensing and Spatial Information Sciences}, vol.~40, pp. 391--395, 2015.

\bibitem[Camps-Valls et~al.(2011)Camps-Valls, Tuia, G{\'o}mez-Chova,
  Jim{\'e}nez, and Malo]{camps2011remote}
G.~Camps-Valls, D.~Tuia, L.~G{\'o}mez-Chova, S.~Jim{\'e}nez, and J.~Malo,
  ``Remote sensing image processing,'' 2011.

\bibitem[Carbone et~al.(2024{\natexlab{a}})Carbone, Restaino, Vivone, and
  Chanussot]{carbone2024model}
A.~Carbone, R.~Restaino, G.~Vivone, and J.~Chanussot, ``Model-based
  super-resolution for sentinel-5p data,'' \emph{IEEE Transactions on
  Geoscience and Remote Sensing}, 2024.

\bibitem[Carbone et~al.(2024{\natexlab{b}})Carbone, Restaino, and
  Vivone]{carbone2024efficient}
A.~Carbone, R.~Restaino, and G.~Vivone, ``Efficient hyperspectral
  super-resolution of sentinel-5p data via dynamic multi-directional cascade
  fine-tuning,'' \emph{IEEE Geoscience and Remote Sensing Letters}, 2024.

\bibitem[Su et~al.(2024)Su, Li, Xu, Fu, and Liu]{su2024review}
H.~Su, Y.~Li, Y.~Xu, X.~Fu, and S.~Liu, ``A review of deep-learning-based
  super-resolution: From methods to applications,'' \emph{Pattern Recognition},
  p. 110935, 2024.

\bibitem[Yang et~al.(2019)Yang, Zhang, Tian, Wang, Xue, and Liao]{yang2019deep}
W.~Yang, X.~Zhang, Y.~Tian, W.~Wang, J.-H. Xue, and Q.~Liao, ``Deep learning
  for single image super-resolution: A brief review,'' \emph{IEEE Transactions
  on Multimedia}, vol.~21, no.~12, pp. 3106--3121, 2019.

\bibitem[Ye et~al.(2023)Ye, Zhao, Hu, and Xie]{ye2023single}
S.~Ye, S.~Zhao, Y.~Hu, and C.~Xie, ``Single-image super-resolution challenges:
  a brief review,'' \emph{Electronics}, vol.~12, no.~13, p. 2975, 2023.

\bibitem[Bashir et~al.(2021)Bashir, Wang, Khan, and
  Niu]{bashir2021comprehensive}
S.~M.~A. Bashir, Y.~Wang, M.~Khan, and Y.~Niu, ``A comprehensive review of deep
  learning-based single image super-resolution,'' \emph{PeerJ Computer
  Science}, vol.~7, p. e621, 2021.

\bibitem[Aburaed et~al.(2023)Aburaed, Alkhatib, Marshall, Zabalza, and
  Al~Ahmad]{aburaed2023review}
N.~Aburaed, M.~Q. Alkhatib, S.~Marshall, J.~Zabalza, and H.~Al~Ahmad, ``A
  review of spatial enhancement of hyperspectral remote sensing imaging
  techniques,'' \emph{IEEE Journal of Selected Topics in Applied Earth
  Observations and Remote Sensing}, vol.~16, pp. 2275--2300, 2023.

\bibitem[Keys(1981)]{keys1981cubic}
R.~Keys, ``Cubic convolution interpolation for digital image processing,''
  \emph{IEEE transactions on acoustics, speech, and signal processing},
  vol.~29, no.~6, pp. 1153--1160, 1981.

\bibitem[Aiazzi et~al.(2013)Aiazzi, Baronti, Selva, and Alparone]{aiazzi2013bi}
B.~Aiazzi, S.~Baronti, M.~Selva, and L.~Alparone, ``Bi-cubic interpolation for
  shift-free pan-sharpening,'' \emph{ISPRS journal of photogrammetry and remote
  sensing}, vol.~86, pp. 65--76, 2013.

\bibitem[Li and Orchard(2001)]{li2001new}
X.~Li and M.~T. Orchard, ``New edge-directed interpolation,'' \emph{IEEE
  transactions on image processing}, vol.~10, no.~10, pp. 1521--1527, 2001.

\bibitem[Hung and Siu(2011)]{hung2011robust}
K.-W. Hung and W.-C. Siu, ``Robust soft-decision interpolation using weighted
  least squares,'' \emph{IEEE Transactions on Image Processing}, vol.~21,
  no.~3, pp. 1061--1069, 2011.

\bibitem[Hung and Siu(2012)]{hung2012fast}
Hung and Siu, ``Fast image interpolation using the bilateral filter,''
  \emph{IET Image Processing}, vol.~6, no.~7, pp. 877--890, 2012.

\bibitem[Schultz and Stevenson(1996)]{schultz1996extraction}
R.~R. Schultz and R.~L. Stevenson, ``Extraction of high-resolution frames from
  video sequences,'' \emph{IEEE transactions on image processing}, vol.~5,
  no.~6, pp. 996--1011, 1996.

\bibitem[Ng et~al.(2007{\natexlab{a}})Ng, Shen, Chaudhuri, and Yau]{ng2007zoom}
M.~K. Ng, H.~Shen, S.~Chaudhuri, and A.~C. Yau, ``Zoom-based super-resolution
  reconstruction approach using prior total variation,'' \emph{Optical
  Engineering}, vol.~46, no.~12, pp. 127\,003--127\,003, 2007.

\bibitem[Fan et~al.(2017)Fan, Wu, Li, and Ma]{fan2017projections}
C.~Fan, C.~Wu, G.~Li, and J.~Ma, ``Projections onto convex sets
  super-resolution reconstruction based on point spread function estimation of
  low-resolution remote sensing images,'' \emph{Sensors}, vol.~17, no.~2, p.
  362, 2017.

\bibitem[Ng et~al.(2007{\natexlab{b}})Ng, Shen, Lam, and Zhang]{ng2007total}
M.~K. Ng, H.~Shen, E.~Y. Lam, and L.~Zhang, ``A total variation regularization
  based super-resolution reconstruction algorithm for digital video,''
  \emph{EURASIP Journal on Advances in Signal Processing}, vol. 2007, pp.
  1--16, 2007.

\bibitem[Sun et~al.(2008)Sun, Xu, and Shum]{sun2008image}
J.~Sun, Z.~Xu, and H.-Y. Shum, ``Image super-resolution using gradient profile
  prior,'' in \emph{2008 IEEE conference on computer vision and pattern
  recognition}.\hskip 1em plus 0.5em minus 0.4em\relax IEEE, 2008, pp. 1--8.

\bibitem[Xiong et~al.(2017)Xiong, Shi, Li, Wang, Liu, and Wu]{xiong2017hscnn}
Z.~Xiong, Z.~Shi, H.~Li, L.~Wang, D.~Liu, and F.~Wu, ``Hscnn: Cnn-based
  hyperspectral image recovery from spectrally undersampled projections,'' in
  \emph{Proceedings of the IEEE international conference on computer vision
  workshops}, 2017, pp. 518--525.

\bibitem[Can and Timofte(2018)]{can2018efficient}
Y.~B. Can and R.~Timofte, ``An efficient cnn for spectral reconstruction from
  rgb images,'' \emph{arXiv preprint arXiv:1804.04647}, 2018.

\bibitem[Han et~al.(2018{\natexlab{a}})Han, Shi, and Zheng]{han2018residual}
X.-H. Han, B.~Shi, and Y.~Zheng, ``Residual hsrcnn: Residual hyper-spectral
  reconstruction cnn from an rgb image,'' in \emph{2018 24th International
  Conference on Pattern Recognition (ICPR)}.\hskip 1em plus 0.5em minus
  0.4em\relax IEEE, 2018, pp. 2664--2669.

\bibitem[Kaya et~al.(2019)Kaya, Can, and Timofte]{kaya2019towards}
B.~Kaya, Y.~B. Can, and R.~Timofte, ``Towards spectral estimation from a single
  rgb image in the wild,'' in \emph{2019 IEEE/CVF International Conference on
  Computer Vision Workshop (ICCVW)}.\hskip 1em plus 0.5em minus 0.4em\relax
  IEEE, 2019, pp. 3546--3555.

\bibitem[Hang et~al.(2021)Hang, Liu, and Li]{hang2021spectral}
R.~Hang, Q.~Liu, and Z.~Li, ``Spectral super-resolution network guided by
  intrinsic properties of hyperspectral imagery,'' \emph{IEEE Transactions on
  Image Processing}, vol.~30, pp. 7256--7265, 2021.

\bibitem[Nathan et~al.(2020)Nathan, Uma, Vinothini, Bama, and
  Roomi]{nathan2020light}
D.~S. Nathan, K.~Uma, D.~S. Vinothini, B.~S. Bama, and S.~Roomi, ``Light weight
  residual dense attention net for spectral reconstruction from rgb images,''
  \emph{arXiv preprint arXiv:2004.06930}, 2020.

\bibitem[Li et~al.(2022)Li, Du, Song, Wu, Li, and Du]{li2022hasic}
J.~Li, S.~Du, R.~Song, C.~Wu, Y.~Li, and Q.~Du, ``Hasic-net: Hybrid attentional
  convolutional neural network with structure information consistency for
  spectral super-resolution of rgb images,'' \emph{IEEE Transactions on
  Geoscience and Remote Sensing}, vol.~60, pp. 1--15, 2022.

\bibitem[Gewali et~al.(2019)Gewali, Monteiro, and Saber]{gewali2019spectral}
U.~B. Gewali, S.~T. Monteiro, and E.~Saber, ``Spectral super-resolution with
  optimized bands,'' \emph{Remote Sensing}, vol.~11, no.~14, p. 1648, 2019.

\bibitem[Wang et~al.(2018)Wang, Yu, Wu, Gu, Liu, Dong, Qiao, and
  Change~Loy]{wang2018esrgan}
X.~Wang, K.~Yu, S.~Wu, J.~Gu, Y.~Liu, C.~Dong, Y.~Qiao, and C.~Change~Loy,
  ``Esrgan: Enhanced super-resolution generative adversarial networks,'' in
  \emph{Proceedings of the European conference on computer vision (ECCV)
  workshops}, 2018, pp. 0--0.

\bibitem[Haris et~al.(2018)Haris, Shakhnarovich, and Ukita]{haris2018deep}
M.~Haris, G.~Shakhnarovich, and N.~Ukita, ``Deep back-projection networks for
  super-resolution,'' in \emph{Proceedings of the IEEE conference on computer
  vision and pattern recognition}, 2018, pp. 1664--1673.

\bibitem[Zhang et~al.(2018{\natexlab{a}})Zhang, Tian, Kong, Zhong, and
  Fu]{zhang2018residual}
Y.~Zhang, Y.~Tian, Y.~Kong, B.~Zhong, and Y.~Fu, ``Residual dense network for
  image super-resolution,'' in \emph{Proceedings of the IEEE conference on
  computer vision and pattern recognition}, 2018, pp. 2472--2481.

\bibitem[Tong et~al.(2017)Tong, Li, Liu, and Gao]{tong2017image}
T.~Tong, G.~Li, X.~Liu, and Q.~Gao, ``Image super-resolution using dense skip
  connections,'' in \emph{Proceedings of the IEEE international conference on
  computer vision}, 2017, pp. 4799--4807.

\bibitem[Li et~al.(2020{\natexlab{a}})Li, Wu, Song, Li, and
  Liu]{li2020adaptive}
J.~Li, C.~Wu, R.~Song, Y.~Li, and F.~Liu, ``Adaptive weighted attention network
  with camera spectral sensitivity prior for spectral reconstruction from rgb
  images,'' in \emph{Proceedings of the IEEE/CVF Conference on Computer Vision
  and Pattern Recognition Workshops}, 2020, pp. 462--463.

\bibitem[Li et~al.(2020{\natexlab{b}})Li, Wu, Song, Xie, Ge, Li, and
  Li]{li2020hybrid}
J.~Li, C.~Wu, R.~Song, W.~Xie, C.~Ge, B.~Li, and Y.~Li, ``Hybrid 2-d--3-d deep
  residual attentional network with structure tensor constraints for spectral
  super-resolution of rgb images,'' \emph{IEEE Transactions on Geoscience and
  Remote Sensing}, vol.~59, no.~3, pp. 2321--2335, 2020.

\bibitem[Zheng et~al.(2021)Zheng, Chen, and Lu]{zheng2021spectral}
X.~Zheng, W.~Chen, and X.~Lu, ``Spectral super-resolution of multispectral
  images using spatial--spectral residual attention network,'' \emph{IEEE
  Transactions on Geoscience and Remote Sensing}, vol.~60, pp. 1--14, 2021.

\bibitem[Zhu et~al.(2021)Zhu, Liu, Hou, Zeng, and Zhang]{zhu2021semantic}
Z.~Zhu, H.~Liu, J.~Hou, H.~Zeng, and Q.~Zhang, ``Semantic-embedded unsupervised
  spectral reconstruction from single rgb images in the wild,'' in
  \emph{Proceedings of the IEEE/CVF International Conference on Computer
  Vision}, 2021, pp. 2279--2288.

\bibitem[Wang et~al.(2023)Wang, Sun, Chehri, and Song]{wang2023review}
X.~Wang, L.~Sun, A.~Chehri, and Y.~Song, ``A review of gan-based
  super-resolution reconstruction for optical remote sensing images,''
  \emph{Remote Sensing}, vol.~15, no.~20, p. 5062, 2023.

\bibitem[Bulat et~al.(2018)Bulat, Yang, and Tzimiropoulos]{bulat2018learn}
A.~Bulat, J.~Yang, and G.~Tzimiropoulos, ``To learn image super-resolution, use
  a gan to learn how to do image degradation first,'' in \emph{Proceedings of
  the European conference on computer vision (ECCV)}, 2018, pp. 185--200.

\bibitem[Zhang and Ling(2020)]{zhang2020supervised}
M.~Zhang and Q.~Ling, ``Supervised pixel-wise gan for face super-resolution,''
  \emph{IEEE Transactions on Multimedia}, vol.~23, pp. 1938--1950, 2020.

\bibitem[Saharia et~al.(2023)Saharia, Ho, Chan, Salimans, Fleet, and
  Norouzi]{Saharia_etal_SR3_image_super-resolution_via_iterative_refinement_PAMI2021}
C.~Saharia, J.~Ho, W.~Chan, T.~Salimans, D.~J. Fleet, and M.~Norouzi, ``Image
  super-resolution via iterative refinement,'' \emph{IEEE Transactions on
  Pattern Analysis and Machine Intelligence}, vol.~45, no.~4, pp. 4713--4726,
  2023.

\bibitem[Han et~al.(2018{\natexlab{b}})Han, Yu, Xue, and Sun]{han2018spectral}
X.~Han, J.~Yu, J.-H. Xue, and W.~Sun, ``Spectral super-resolution for rgb
  images using class-based bp neural networks,'' in \emph{2018 Digital Image
  Computing: Techniques and Applications (DICTA)}.\hskip 1em plus 0.5em minus
  0.4em\relax IEEE, 2018, pp. 1--7.

\bibitem[Mei et~al.(2022)Mei, Geng, Hou, and Du]{mei2022learning}
S.~Mei, Y.~Geng, J.~Hou, and Q.~Du, ``Learning hyperspectral images from rgb
  images via a coarse-to-fine cnn,'' \emph{Science China Information Sciences},
  vol.~65, pp. 1--14, 2022.

\bibitem[Guarino et~al.(2023)Guarino, Ciotola, Vivone, and
  Scarpa]{guarino2023band}
G.~Guarino, M.~Ciotola, G.~Vivone, and G.~Scarpa, ``Band-wise hyperspectral
  image pansharpening using cnn model propagation,'' \emph{IEEE Transactions on
  Geoscience and Remote Sensing}, 2023.

\bibitem[Qu et~al.(2021)Qu, Hou, Dong, Xiao, Du, and Li]{qu2021dual}
J.~Qu, S.~Hou, W.~Dong, S.~Xiao, Q.~Du, and Y.~Li, ``A dual-branch detail
  extraction network for hyperspectral pansharpening,'' \emph{IEEE Transactions
  on Geoscience and Remote Sensing}, vol.~60, pp. 1--13, 2021.

\bibitem[Hidalgo et~al.(2021)Hidalgo, Cort'es, and
  Bravo]{hidalgo2021dimensionality}
D.~R. Hidalgo, B.~B. Cort'es, and E.~C. Bravo, ``Dimensionality reduction of
  hyperspectral images of vegetation and crops based on self-organized maps,''
  \emph{Information Processing in Agriculture}, vol.~8, no.~2, pp. 310--327,
  2021.

\bibitem[He et~al.(2023)He, Yuan, Li, Xiao, Liu, Shen, and
  Zhang]{he2023spectral}
J.~He, Q.~Yuan, J.~Li, Y.~Xiao, D.~Liu, H.~Shen, and L.~Zhang, ``Spectral
  super-resolution meets deep learning: Achievements and challenges,''
  \emph{Information Fusion}, vol.~97, p. 101812, 2023.

\bibitem[Sifre(2014)]{sifre2014rigidmotion}
L.~Sifre, ``Rigid-motion scattering for image classification,'' PhD thesis,
  Ecole Polytechnique - CMAP, 2014.

\bibitem[Hung et~al.(2019)Hung, Zhang, and Jiang]{hung2019real}
K.-W. Hung, Z.~Zhang, and J.~Jiang, ``Real-time image super-resolution using
  recursive depthwise separable convolution network,'' \emph{IEEE Access},
  vol.~7, pp. 99\,804--99\,816, 2019.

\bibitem[Hussain and Lall(2023)]{hussain2023depth}
S.~Hussain and B.~Lall, ``Depth separable-cnn for improved spectral
  super-resolution,'' \emph{IEEE Access}, vol.~11, pp. 23\,063--23\,072, 2023.

\bibitem[Sun et~al.(2024)Sun, Ke, Liu, Lu, Li, Yang, and
  Zhang]{sun2024enhanced}
W.~Sun, R.~Ke, Z.~Liu, H.~Lu, D.~Li, F.~Yang, and L.~Zhang, ``Enhanced feature
  refinement network based on depthwise separable convolution for lightweight
  image super-resolution,'' \emph{Symmetry}, vol.~16, no.~11, p. 1406, 2024.

\bibitem[Muhammad et~al.(2021)Muhammad, Aramvith, and Onoye]{muhammad2021multi}
W.~Muhammad, S.~Aramvith, and T.~Onoye, ``Multi-scale xception based depthwise
  separable convolution for single image super-resolution,'' \emph{Plos one},
  vol.~16, no.~8, p. e0249278, 2021.

\bibitem[Jiang et~al.(2020)Jiang, Huang, and Hu]{jiang2020single}
Z.~Jiang, Y.~Huang, and L.~Hu, ``Single image super-resolution: Depthwise
  separable convolution super-resolution generative adversarial network,''
  \emph{Applied Sciences}, vol.~10, no.~1, p. 375, 2020.

\bibitem[Chollet(2017)]{chollet2017xception}
F.~Chollet, ``Xception: Deep learning with depthwise separable convolutions,''
  in \emph{Proceedings of the IEEE conference on computer vision and pattern
  recognition}, 2017, pp. 1251--1258.

\bibitem[Junejo and Ahmed(2021)]{junejo2021depthwise}
I.~N. Junejo and N.~Ahmed, ``Depthwise separable convolutional neural networks
  for pedestrian attribute recognition,'' \emph{SN Computer Science}, vol.~2,
  no.~2, p. 100, 2021.

\bibitem[{Copernicus Data Space}(2024)]{CopernicusDataSpace}
\BIBentryALTinterwordspacing
{Copernicus Data Space}, ``{Copernicus Data Space Ecosystem},'' 2024, accessed:
  2024-12-23. [Online]. Available: \url{https://dataspace.copernicus.eu/}
\BIBentrySTDinterwordspacing

\bibitem[Merry et~al.(2024)Merry, Ossorgin, and Mekus]{merry2024backbone}
K.~Merry, J.~Ossorgin, and Z.~Mekus, ``Backbone architectures for space domain
  awareness,'' in \emph{Proceedings of the< a href= https://www. amostech. com>
  Advanced Maui Optical and Space Surveillance (AMOS) Technologies
  Conference</a}, 2024, p.~16.

\bibitem[Nakanishi et~al.(2024)Nakanishi, Kubo, Yasui, and
  Ishii]{nakanishi2024robust}
K.~Nakanishi, A.~Kubo, Y.~Yasui, and S.~Ishii, ``Robust off-policy
  reinforcement learning via soft constrained adversary,'' \emph{arXiv preprint
  arXiv:2409.00418}, 2024.

\bibitem[Wang et~al.(2004)Wang, Bovik, Sheikh, and Simoncelli]{wang2004image}
Z.~Wang, A.~C. Bovik, H.~R. Sheikh, and E.~P. Simoncelli, ``Image quality
  assessment: from error visibility to structural similarity,'' \emph{IEEE
  transactions on image processing}, vol.~13, no.~4, pp. 600--612, 2004.

\bibitem[An et~al.(2014)An, Gong, Yin, Wang, Pan, Zhang, Lu, Yang, Toth,
  Schiessl, et~al.]{an2014orientation}
X.~An, H.~Gong, J.~Yin, X.~Wang, Y.~Pan, X.~Zhang, Y.~Lu, Y.~Yang, Z.~Toth,
  I.~Schiessl \emph{et~al.}, ``Orientation-cue invariant population responses
  to contrast-modulated and phase-reversed contour stimuli in macaque v1 and
  v2,'' \emph{PLoS One}, vol.~9, no.~9, p. e106753, 2014.

\bibitem[Zhang et~al.(2018{\natexlab{b}})Zhang, Isola, Efros, Shechtman, and
  Wang]{zhang2018unreasonable}
R.~Zhang, P.~Isola, A.~A. Efros, E.~Shechtman, and O.~Wang, ``The unreasonable
  effectiveness of deep features as a perceptual metric,'' in \emph{Proceedings
  of the IEEE conference on computer vision and pattern recognition}, 2018, pp.
  586--595.

\bibitem[Stenberg et~al.(2010)Stenberg, Rossel, Mouazen, and
  Wetterlind]{stenberg2010visible}
B.~Stenberg, R.~A.~V. Rossel, A.~M. Mouazen, and J.~Wetterlind, ``Visible and
  near infrared spectroscopy in soil science,'' \emph{Advances in agronomy},
  vol. 107, pp. 163--215, 2010.

\end{thebibliography}

\end{document}